\documentclass[pra,aps,showpacs,twocolumn]{revtex4}
\usepackage{graphics,bm}
\usepackage{amssymb,ulem}
\usepackage{amsmath}
\usepackage{epsfig}
\usepackage[usenames]{color}

\begin{document}

\title{Interaction effects on dynamic correlations in non-condensed Bose gases}

\author{A. Bezett}
\affiliation{Institute for Theoretical Physics, Utrecht
University, Leuvenlaan 4, 3584 CE Utrecht, The Netherlands}

\author{H.J. van Driel}
\affiliation{Institute for Theoretical Physics, Utrecht
University, Leuvenlaan 4, 3584 CE Utrecht, The Netherlands}

\author{M.P. Mink}
\affiliation{Institute for Theoretical Physics, Utrecht
University, Leuvenlaan 4, 3584 CE Utrecht, The Netherlands}

\author{H.T.C. Stoof}
\affiliation{Institute for Theoretical Physics, Utrecht
University, Leuvenlaan 4, 3584 CE Utrecht, The Netherlands}

\author{R.A. Duine}
\affiliation{Institute for Theoretical Physics, Utrecht
University, Leuvenlaan 4, 3584 CE Utrecht, The Netherlands}

\date{\today}

\begin{abstract}
We consider dynamic, i.e., frequency-dependent, correlations in non-condensed ultracold atomic Bose gases. In particular, we consider the single-particle correlation function and its power spectrum. We compute this power spectrum for a one-component Bose gas, and show how it depends on the interatomic interactions that lead to a finite single-particle relaxation time. As another example, we consider the power spectrum of spin-current fluctuations for a two-component Bose gas and show how it is determined by the spin-transport relaxation time.
\end{abstract}


\maketitle

\newcommand{\etal}{\emph{et al.}}
\newcommand{\rC}{\text{\bf{C}}} 
\newcommand{\rI}{\text{\bf{I}}} 
\newcommand{\PC}{\mathcal{P}_{\rC}}
\newcommand{\bef}{\hat{\psi}}
\newcommand{\bfI}{\bef_{\rI}}
\newcommand{\cf}{\psi_{\rC}}
\newcommand{\cfp}{\psi_{\rC'}}
\newcommand{\ecut}{\epsilon_{\rm cut}}
\newcommand{\CF}{c-field}
\newcommand{\Nc}{N_{\rm{cond}}}
\newcommand{\psic}{\psi_{\rm{cond}}}
\newcommand{\ac}{\alpha_{\rm{cond}}}
\newcommand{\thold}{t_{\rm{hold}}}
\newcommand{\nmin}{n_{\min}}
\def\x{\mathbf{r}}
\newcommand{\xa}{(\x)}
\def\s{{\rm s}}
\def\bgam{{\bm \Gamma}}
\def\v{{\bm v}}
\def\F{{\bm F}}
\def\bX{{\bm X}}
\def\bx{{\bm x}}
\def\bk{{\bm k}}
\def\bK{{\bm K}}
\def\bq{{\bm q}}
\def\br{{\bm r}}
\def\bp{{\bm p}}
\def\half{\frac{1}{2}}
\def\args{(\bm, t)}
\def\x{{\bm x}}
\def\k{{\bm k}}
\def\bnab{{\bm \nabla}}
\def\J{{\rm J}}
\def\j{{\bf j}}
\section{Introduction}
Since the early days of the field, the measurement techniques used to characterize cold-atom systems have developed significantly. Initially, the most common experimental probe to address cold atomic vapors consisted of absorption imaging of an expanding atomic cloud, thereby providing information on its equilibrium velocity distribution only. Measurements on solid-state systems, however, are typically in the linear-response regime and measure the change of an observable in response to a spatial and/or temporal periodic external perturbation. These measurements are therefore typically characterized by the frequency and momentum dependence of equilibrium correlation functions, and in solid-state physics a large number of experimental techniques have been developed over the past two centuries to probe different domains in frequency and momentum. Fortunately, in several recent developments, the field of cold atoms is rapidly catching up in the number of available experiment probes which are sensitive to momentum and frequency. 

In this article we focus on dynamic correlations, i.e., the frequency or time dependence of equilibrium correlation functions. For cold-atom systems the most prominent techniques to probe these are Bragg spectroscopy \cite{stenger1999,stamperkurn1999,papp2008,veeravalli2008}, radio-frequency (RF) spectroscopy \cite{gupta2003,chin2004,stewart2008}, and, most recently, impact ionization via a scanning electron microscope \cite{guarrera2011}. These techniques respectively probe the structure factor, the spectral function, and the second-order temporal correlations. Most research considers either partially Bose-Einstein condensed gases \cite{stenger1999,stamperkurn1999,papp2008} or Fermi gases in the crossover regime \cite{veeravalli2008,gupta2003,chin2004,stewart2008}. Here we focus on the non-condensed regime of a cold gas of bosonic atoms and on how interactions influence the dynamic correlations in this case.

In particular, in Sec.~\ref{sec:singleparticle} we consider the frequency dependence of the scattering rate as measured in an RF-spectroscopy experiment. We compute this scattering rate and find that interactions significantly change the spectrum with respect to the non-interacting case, as they lead to a finite lifetime of the quasi-particles. We also point out that this scattering rate, when measured above but close to the critical temperature for Bose-Einstein condensation, contains information on the critical exponents of the transition. At the end of this section we assess the importance of the trapping potential with numerical simulations.

As another example of dynamical correlations, we consider, in Sec.~\ref{sec:spincurrentflucs}, spin-current fluctuations in a two-component Bose gas. These could be measured by a spin-resolved generalization of the above-mentioned scanning-electron microscope techniques. The power spectrum of the spin-current fluctuations contains, via the fluctuation-dissipation theorem, information on the spin resistivity of the gas that is fully determined by interactions in this system. Moreover, this provides an example of how an equilibrium measurement can be used to determine a transport coefficient when steady-state transport measurements are not readily available as is typically the case for trapped atomic gases. We end in Sec.~\ref{sec:discussion} with our conclusions, and a brief discussion and outlook.

\section{Single-particle correlations} \label{sec:singleparticle}
In this section we consider dynamic, i.e., temporal, correlations contained in the single-particle Green's function as measured in an RF-spectroscopy experiment. In the first part we give general theoretical considerations that relate the single-particle correlations to the spectral function. Hereafter, we consider the non-interacting and interacting case separately, and also present results from projected Gross-Pitaevskii equation simulations.

\subsection{Theoretical framework}
In the RF-spectroscopy experiments by Stewart {\it et al.} \cite{stewart2008} one measures the momentum and frequency-dependent scattering rate $R(\bq, \omega)$, determined by
\begin{eqnarray}
\label{eq:intensitygeneral}
 R (\bq, \omega) &\propto& \int d (t-t') \int d\bX \int d\bx e^{-i\bq\cdot \bx +i\omega (t-t')} \nonumber \\ && \times \langle \hat \psi^\dagger (\bX+\bx/2,t) \hat \psi (\bX-\bx/2,t')\rangle~,
\end{eqnarray}
and thus determined by the single-particle correlation function $\langle \hat \psi^\dagger (\bx,t) \hat \psi (\bx',t')\rangle$, where $\langle \cdots \rangle$ is an equilibrium expectation value. Here,  $\hat \psi (\bx,t)$ is the Heisenberg annihilation field operator at position $\bx$ and time $t$. Furthermore, the integral over position $\bX$ in Eq.~(\ref{eq:intensitygeneral}) is an average over the inhomogeneous density of the gas.

The above scattering rate can be worked out as follows. First we note that the so-called ``lesser"($<$) and ``greater" ($>$) Green's functions, are defined by
\begin{eqnarray}
\label{eq:defglesserandgreater}
  \pm i G^< (\bx,t;\bx',t') &=& \left\langle \hat \psi^\dagger (\bx',t') \hat \psi (\bx,t) \right\rangle~; \nonumber \\
  i G^>(\bx,t;\bx',t') &=& \left\langle \hat \psi (\bx,t) \hat \psi^\dagger (\bx',t') \right\rangle~.
\end{eqnarray}
These are determined by the retarded Green's function
\begin{eqnarray}
 && i G^< (\bx,\bx'; \omega) = \mp 2 n_{B/F} (\hbar \omega-\mu) {\rm Im} \left[ G^{(+)} (\bx,\bx';\omega) \right]~; \nonumber \\
  && i G^> (\bx,\bx'; \omega) =- 2\left[1\pm n_{B/F} (\hbar \omega-\mu) \right]\nonumber \\
  && \times {\rm Im} \left[ G^{(+)} (\bx,\bx';\omega) \right]~,
\end{eqnarray}
via the fluctuation-dissipation theorem. Here, the spectral function is given by
\begin{equation}
\label{eq:defspectrfct}
  \rho (\bx,\bx';\omega) = - \frac{1}{\pi\hbar} {\rm Im} \left[ G^{(+)} (\bx,\bx';\omega) \right]~,
\end{equation}
in terms of the imaginary part of the temporal Fourier transform of the retarded Green's function
\begin{equation}
\label{eq:retardedgfdef}
  G^{(+)} (\bx,t;\bx',t') = i \theta (t-t') \left\langle \left[\hat \psi (\bx,t),\hat \psi^\dagger (\bx,t) \right]\right\rangle~.
\end{equation}
Furthermore, the Bose/Fermi function $n_{B/F} (\hbar\omega) = (e^{\beta \hbar \omega}\mp 1)^{-1}$ where $\beta=1/k_BT$ is the inverse thermal energy. The chemical potential is denoted by $\mu$. In all of the above, upper (lower) signs refer to bosons (fermions). From now on, however, we consider only bosons.

The position dependence of the
spectral function and the Green's function can usually be treated in the local-density approximation. In this case the retarded Green's function is given
by
\begin{eqnarray}
\label{eq:retardedgflocaldensity}
  && G^{(+)} (\bx,\bx'; \omega ) = \int \frac{d\bk}{(2 \pi)^3}
  e^{i\bk\cdot (\bx-\bx')} \nonumber \\
  && \times \frac{\hbar}{\hbar \omega^+ -\epsilon_\bk - V_{\rm ext} \left( \frac{\bx+\bx'}{2}\right) - \hbar \Sigma^{(+)} \left( \bk,
  \frac{\bx+\bx'}{2},\omega\right)}~, \nonumber \\
\end{eqnarray}
where the retarded self-energy $\hbar \Sigma^{(+)} \left( \bk,\bX,\omega\right)$ is determined by evaluating (in a suitable approximation) the retarded self-energy for a homogeneous system
 and then replacing $\mu \rightarrow \mu - V_{\rm ext} \left( \bX \right)$. As a result, the above Green's function follows from the retarded
 Green's function for a homogeneous system by using this Green's function and adding the confining potential $V_{\rm ext} (\bX)$ to the single-particle energy. In the above the dispersion is $\epsilon_\bk =\hbar^2 \bk^2/2m$ with $m$ the mass of a single atom. We now introduce the Fourier transform of the spectral function to relative position and momentum variables via
\begin{equation}
\label{eq:rhorelvars}
 \rho (\bk, \bX, \omega) = \int d\bx e^{-i\bk\cdot\bx} \rho (\bX+\bx/2,\bX-\bx/2;\omega)~,
\end{equation}
which yields the scattering rate
\begin{equation}
\label{eq:powerspectrumlocaldens}
  R(\bq, \omega) \propto 2\pi \hbar \int d\bX \rho (\bq,\bX,\omega) n_B (\hbar \omega-\mu)~.
\end{equation}

\subsection{Non-interacting case} We first consider the non-interacting case. In that case the self-energy is zero and the spectral function given by
\begin{equation}
\label{eq:spectrfctnonint}
  \rho_0 (\bk,\bX,\omega) = \delta (\hbar \omega - \epsilon_\bk-V_{\rm ext}(\bX))~.
\end{equation}
From this we find the scattering rate
\begin{equation}
\label{eq:scatteringrateideal}
 R_0(\bq, \omega) \propto \frac{8\sqrt{2}\pi^2  \hbar}{m^{3/2} \omega_x\omega_y\omega_z} \theta (\hbar \omega -\epsilon_\bq) n_B (\hbar \omega-\mu)\sqrt {\hbar \omega -\epsilon_\bq } ~,
\end{equation}
where we have taken
\begin{equation} \label{eq:vext}
V_{\rm ext} (\bx) = m(\omega^2_x x^2 + \omega^2_y x^2+ \omega^2_z z^2)/2~.
\end{equation}
Within this ideal-gas approximation, the chemical potential is determined by solving the equation for the total number of particles $N$, i.e.,
\begin{equation}
\label{eq:numbereqidealgas}
  N = \int d\bX\int \frac{d\bk}{(2\pi)^3} \int d (\hbar \omega) n_B (\hbar \omega - \mu) \rho_0 (\bk, \bX, \omega)~.
\end{equation}
The Heaviside step function $\theta$, the Bose function, and the factor $\sqrt {\hbar \omega -\epsilon_\bq }$ in Eq.~(\ref{eq:scatteringrateideal}) respectively reflect the threshold for obeying energy conservation, the fact that the scattering rate depends on the number of available particles at the scattering energy, and the modification of the local density of states by the trapping potential. Without the latter the scattering rate would be proportional to $\delta (\hbar \omega -\epsilon_\bq )$ rather than this square root.

\subsection{Interactions}

The hamiltonian for the single-component Bose gas is
\begin{eqnarray}
\label{eq:hamspinless}
  \hat H & =&\int d \bx \hat \psi^\dagger (\bx) \left[ - \frac{\hbar^2 \bm{\nabla}^2}{2m} + V_{\rm ext} (\bx) \right]
  \hat \psi (\bx)\nonumber \\
  & +& \frac{1}{2} T^{2B} \int d\bx\hat \psi^\dagger (\bx) \hat \psi^\dagger (\bx) \hat \psi (\bx) \hat \psi (\bx) ~,
\end{eqnarray}
with $T^{2B} = 4 \pi a \hbar^2/m$
the two-body $T$-matrix in terms of the $s$-wave scattering length $a$. The Schr\"odinger creation and annihilation operators for the atoms are $\hat \psi^\dagger (\bx)$
and $\hat \psi (\bx)$.

The presence of interactions changes the spectral function. Instead of the delta function in Eq.~(\ref{eq:spectrfctnonint}), it is now equal to
\begin{multline}
\label{eq:spectrfctint}
\rho (\bk,\bX,\omega) =- \frac{1}{\pi} \\
\frac{\hbar \Im \Sigma(\bk,\bX,\omega)}{[-\hbar \omega + \epsilon_\bk+V_{\rm ext}(\bX)+ \Re \Sigma(\bk,\bX,\omega)]^2+ \hbar^2 \left[\Im \Sigma(\bk,\bX,\omega)\right]^2}.
\end{multline}
To determine the retarded self-energy $\Sigma(\bk,\bX,\omega)$ we restrict ourselves to the homogeneous case so that $\Sigma(\bk,\bX,\omega)= \Sigma(\bk,\omega)$. We adopt the sunset approximation \cite{rakpong}, which is the simplest approximation that yields broadening of the spectral function due to collisions. In this approximation, the imaginary part of the self-energy is given by
\begin{multline} \label{bla}
\hbar \Im \Sigma(\bk,\omega) =  \frac{- (2 \pi)^3\pi (T^{2B})^2}{(2 \pi)^9} \int d \bk_1 d \bk_2  d\bk_3 \\ \delta(\bk +  \bk_2 - \bk_3 - \bk_4 ) \delta(\hbar \omega + \epsilon(\bk_2) - \epsilon(\bk_3) - \epsilon(\bk_4)) \\ [n_B(\bk_2) (1 + n_B(\bk_3)) (1 + n_B(\bk_4)) \\ - (1 + n_B(\bk_2)) n_B(\bk_3) n_B(\bk_4)],
\end{multline}
where the Hartree-Fock mean-field shift $2 T^{2B}n$  (that arises in a first-order approximation to the self-energy) should be included in the single-particle dispersion via the substitution $\mu \to \mu - 2 T^{2B}n$, but is omitted here. Note that in the above we use the shorthand notation $n_B (\bk) \equiv n_B (\epsilon_\bk-\mu)$. We evaluate the above expression for the self-energy numerically to obtain the imaginary part, and then the real part can be obtained by the Kramers-Kronig transform.

\begin{figure}
\begin{center}
\includegraphics[width = 0.45 \textwidth]{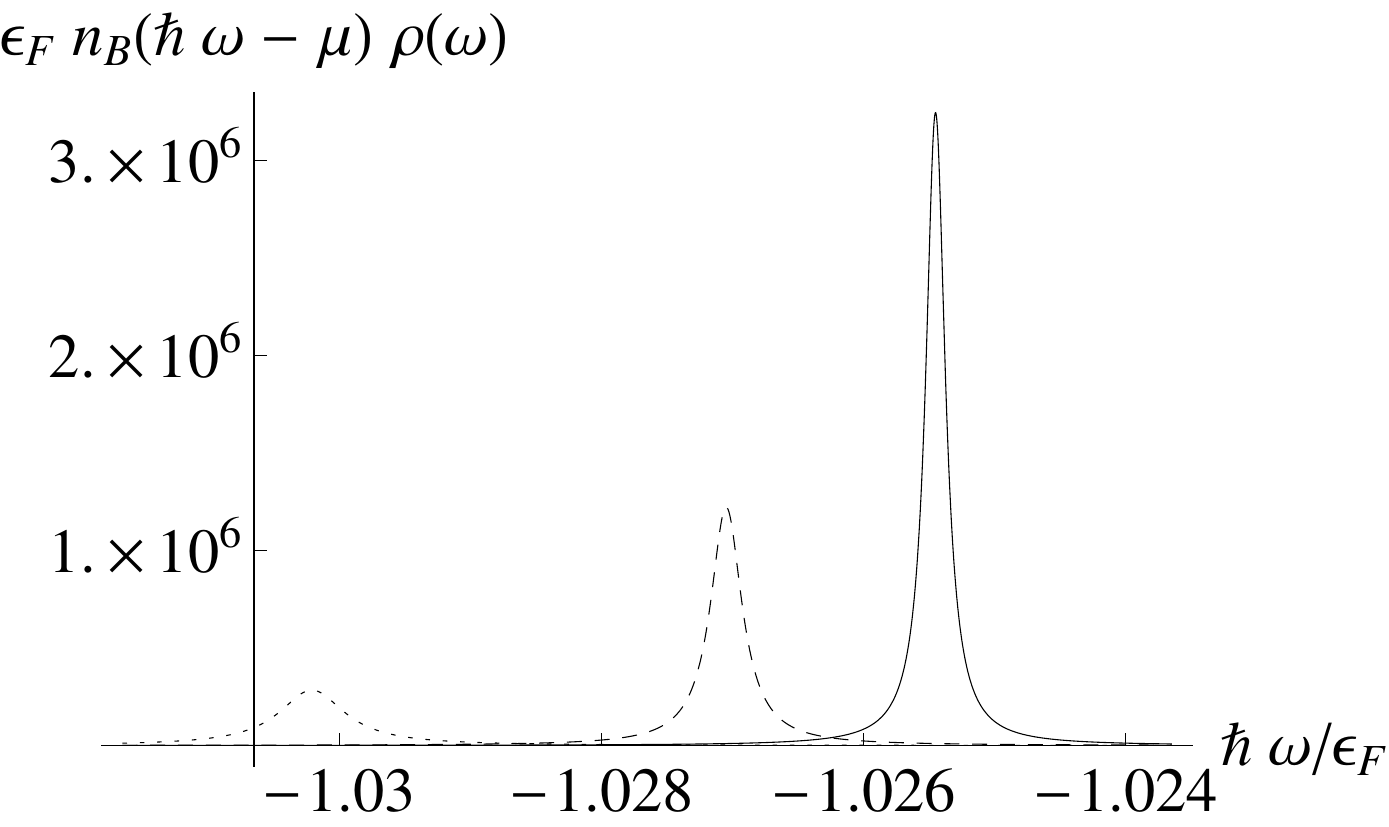}
\caption{\label{fig:nbrho} The dimensionless product $\epsilon_F \rho (\bq=0,\omega) n_B (\hbar \omega - \mu)$ as a function of $\hbar \omega /\epsilon_F$ for dimensionless interaction parameter $k_F a = 1.07$. The peaks correspond to $T/T_c = 1.06$ (dotted curve), $T/T_c = 1.04$ (dashed curve), and $T/T_c = 1.03$ (solid curve).}
\end{center}
\end{figure}

For definiteness we consider zero momentum so that $\bq=0$, and plot the product $\rho (\bq=0,\omega) n_B (\hbar \omega-\mu)$, which determines the scattering rate in a homogenous system for a particular chemical potential, as seen from Eq.~(\ref{eq:powerspectrumlocaldens}). The result is shown in Fig.~\ref{fig:nbrho}, where we show the dimensionless product $\epsilon_F \rho (\bq=0,\omega) n_B (\hbar \omega - \mu)$ as a function of $\hbar \omega /\epsilon_F$ for dimensionless interaction parameter $k_F a = 1.07$. Here, $\epsilon_F = \hbar^2 k_F^2/2 m$ is the Fermi energy corresponding to the particle density $n$, with the Fermi wavenumber $k_F = (6 \pi^2 n)^{1/3}$. Note that the Fermi energy and Fermi momentum introduced here merely serve the purpose of introducing an energy and momentum scale independent of temperature, but have no particular physical significance for the momentum distribution as the system under consideration is bosonic. The peaks in Fig.~\ref{fig:nbrho} correspond to $T/T_c = 1.06$ (dotted curve), $T/T_c = 1.04$ (dashed curve), and $T/T_c = 1.03$ (solid curve), where $T_c$ is the critical temperature for Bose-Einstein condensation. For this particular value of $k_F a$, the $T_c$ in this model is $0.47 T_F$.  The spectral function is peaked when $\hbar \omega$ is equal to the single-particle dispersion. Obtaining a self-consistent dispersion is hard as it requires solving the chemical potential from the equation for the density
\begin{equation}
\label{eq:densityinthomog}
 n = \int \frac{d\bk}{(2\pi)^3} \int d(\hbar \omega) n_B (\hbar \omega - \mu) \rho (\bk,\omega)~,
\end{equation}
of which the right-hand side also depends in principle on density via the above-mentioned mean-field shift. By leaving the chemical potential $\mu$ undetermined, we do not attempt an accurate prediction of the positions of the peaks in Fig.~\ref{fig:nbrho} at given temperature and density. Rather, we focus on the finite quasi-particle lifetimes due to interactions, that are reflected in broadening (with respect to the delta function appropriate for the non-interacting case) of the peaks. From Fig.~\ref{fig:nbrho} we see that upon approaching $T_c$ from above, the peak height increases while the peak width decreases. We consider now the scaling relation of the spectral function that holds in the critical region and is given by \cite{amitbook}
$$
\rho(\lambda \bk, \lambda^z \omega,\lambda^{1/\nu}(T-T_c)) = \frac{1}{\lambda^{2-\eta}} \rho(\bk,\omega,T-T_c).
$$
The prediction for the critical exponents within the sunset approximation are $\nu = 1/2,z=2,\eta=0$. One may use this scaling relation to collapse experimentally obtained spectral functions (allowing also for a shift of the energy axis), and so achieve an experimental result for the exponents.

\subsection{Numerical results}
In this section we determine the single-particle correlation function using simulations based on the projected Gross-Pitaevskii equation (PGPE) formalism. These simulations incorporate both the effects of the trapping potential and interactions.

\subsubsection{PGPE formalism} \label{secPGPEformalism}
The projected Gross-Pitaevskii equation (PGPE) formalism is developed in detail in Ref. \cite{PGPE}, and so we will not repeat it here.
The crux of the method is that the Bose field operator is split into two parts according to
\begin{equation}
\hat\psi(\bx) = \cf(\bx) + \bfI(\bx),\label{EqfieldOp}
\end{equation}
where $\cf$ is the coherent classical field  and $\bfI$ is the incoherent field operator (see Ref.~\cite{cfieldRev2008}).
These fields are defined as the low and high-energy projections of the full quantum-field operator, separated by the energy $\ecut$. In this work, we study only the coherent region, as it has been shown \cite{Bezett2008} that the coherent region contains all the long-range coherences in the system that will be relevant for our study.
The equation of motion for $\cf$ is the PGPE
\begin{eqnarray}
i\hbar\frac{\partial \cf }{\partial t} &=&  \left[- \frac{\hbar^2 \bm{\nabla}^2}{2m} + V_{\rm ext} (\bx) \right]\cf
 \nonumber \\ &&+ \PC\left\{ T^{2B} |\cf|^2\cf\right\}, \label{PGPE}
\end{eqnarray}
where the projection operator
\begin{equation}
\PC\{ F(\bx)\}\equiv\sum_{n\in\rC}\varphi_{n}(\bx)\int
d\bx'\,\varphi_{n}^{*}(\bx') F(\bx'),\label{eq:projectorC}\\
\end{equation}
formalises our basis set restriction of $\cf$ to the coherent region. The main approximation used to arrive at the PGPE is to neglect dynamical couplings to the incoherent region \cite{Davis2001b}.

Eq.~(\ref{PGPE}) is ergodic, so that as $\cf$ evolves in time it samples the equilibrium microstates of the system. Time averaging can therefore be used to obtain macroscopic equilibrium properties. We begin our study by finding equilibrium states of the Bose gas, for temperatures around the critical temperature, following the method of Ref. \cite{Bezett2009a}.

 Typically evolution times of order 20 trap periods are used for the system to relax towards equilibrium \cite{Blakie2008a}, before properties of the equilibrium states are sampled using time-averaging. We typically use around $7000$ samples over 140 trap periods  of our simulation to perform such averages. Details of the calculation of these equilibrium properties can be found in  Refs.~\cite{Bezett2009a,cfieldRev2008} and references therein. In this work, we study the time-dependent behavior of the Bose gas. Ergodic time averaging can obviously not be used when studying the dynamics of the system, so we instead take ensemble averages for dynamic properties to improve the statistics of our results.

We aim to find the single-particle response function, $G^{(+)} (\bx,\bx';\omega)$ of the system. To this end, a perturbation is added to the Hamiltonian of the PGPE, which is now given by
\begin{eqnarray}
i\hbar\frac{\partial \cf }{\partial t} &=&  \left[- \frac{\hbar^2 \bm{\nabla}^2}{2m} + V_{\rm ext} (\bx) \right]\cf
 \nonumber \\ && + \PC\left\{ T^{2B} |\cf|^2\cf\right\} +F(\bx,t)~,
\end{eqnarray}
where
\begin{equation}
F (\bx,t) = F_0 e^{-A\bx^2}e^{-i \omega_0 t}. \label{Perturbation}
\end{equation}
The inclusion of the exponent $-A\bx^2$ restricts the perturbation to a small area within the very center of the trapped system. The full width at half maximum of the perturbation is $0.3\mu m$, while the system width for $\rC$ in the short (tight trap) direction is around $5.5 \mu m$. The amplitude of this perturbation is kept minimal to ensure the system is in the linear-response regime. The single-particle response function, $G^{(+)} (\bx,\bx';t-t^\prime)$ determines the perturbation of the wave function via
\begin{eqnarray}
\langle \cf(\bx,t) \rangle_F &=& \langle \cf(\bx,t) \rangle_0 \\
&+& \frac{1}{\hbar} \int dt^\prime \int d\bx' G^{(+)}(\bx,\bx';t-t^\prime) F(\bx',t^\prime)~. \nonumber
\end{eqnarray}
Making use of the fact that $G^{(+)}(\bx,\bx';t-t^\prime)$ goes to zero as a function of $|\bx-\bx'|$ on a length scale that is much smaller than the width of the perturbation, we conclude that the response is essentially local and of the form
\begin{equation}
\langle \cf(\bx,\omega) \rangle_F - \langle \cf(\bx,\omega) \rangle_0 \propto e^{-A\bx^2} K(\bx,\omega_0) \delta(\omega-\omega_0)~,\label{K_fourier}
\end{equation}
where $\omega_0$ is the perturbation frequency, angled brackets denote ensemble averages, and $\langle \cdots \rangle_F$ indicates such an average in the presence of the perturbation. 

Integrating Eq. (\ref{K_fourier}) further over all frequencies $\omega$ gives the response function $K(\bx,\omega_0)$ in terms of the Fourier transform of the perturbed and unperturbed wave functions. From now on, we focus on $\bx=0$. We take the following steps to find the response function.
\begin{itemize}
\item Find an ensemble of equilibrium states, all with the same approximate temperature and total number.
\item Add the perturbation of Eq. (\ref{Perturbation}) to each of the states, and evolve the PGPE. This is done for a range of frequencies, $\omega$. The PGPE is also evolved without perturbation.
\item The ensemble average of all wave functions is taken, at the very center point of the trap, for the full time length of the simulation.
\item The Fourier transform is taken of the ensemble average, for all frequencies, including the non-perturbed result.
\item The integral over the frequency is taken, to find $K({\bf 0},\omega_0)$, for each $\omega_0$.
\end{itemize}

\subsubsection{Results}

We simulate three systems that span the critical region. These systems are created by fixing $\ecut$, and the number of particles in the coherent region, $N_\rC$, and then choosing three different values for the energy of the coherent region, $E_\rC$, as detailed in Ref. \cite{Bezett2009a}. This process results in three systems that all have different particle number, but which have increasing relative temperature, given by $T/T_c$, where $T_c$ is the critical temperature. The systems have harmonic trap frequencies $\omega_x = 2\pi \times 365$ Hz, $\omega_y = \omega_z = 2\pi \times 129$ Hz. The equilibrium properties for our systems are given in Table \ref{tab:gamma}.

From Ref. \cite{Bezett2009a}, it can be seen that a best estimate for the critical temperature is $0.96T_{c1}$,  where
\begin{eqnarray}
T_{c1} &=& T_{c0} - \left(0.73\frac{\bar{\omega}}{\omega}{N^{-\frac{1}{3}}} + 1.33\frac{a}{a_{\rm{ho}}}N^{\frac{1}{6}}\right)T_{c0}, \label{Tc1}
\end{eqnarray}
with
\begin{equation}
{k_B}T_{c0} = 0.94\hbar\omega N^{1/3},
\end{equation}
and $\omega = (\omega_x \omega_y \omega_z)^{1/3}$, $\bar{\omega} = (\omega_x + \omega_y + \omega_z)/3$, and $a_{\rm{ho}} = \sqrt{\hbar/m\omega}$, see  Ref. \cite{Giorgini1996a}.  The two terms in brackets in Eq. (\ref{Tc1}) correspond to the finite-size ($\propto N^{-1/3}$)  and mean-field interaction  ($\propto N^{1/6}$)  shifts of the critical temperature, respectively. We use $0.96T_{c1}$ as the critical temperature in this paper,  where our relative temperature is defined as $T/(0.96T_{c1})$.

\begin{figure}
\includegraphics[width=3.5in, keepaspectratio]{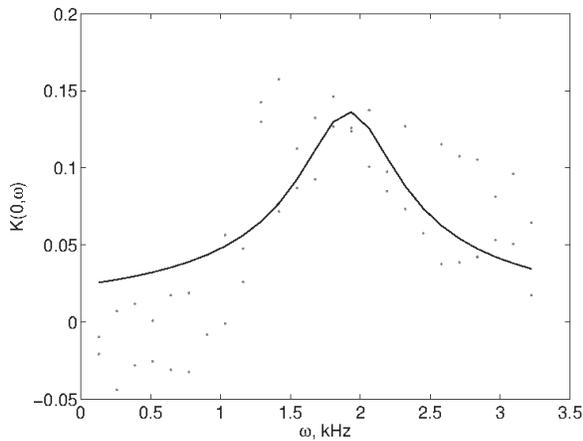}
 \caption{\label{Fig_15} Data points (crosses) and fitted curve (line) for $T = 1.00T_c$ }
\end{figure}

To study the temperature dependence, we fit the function

\begin{equation}
 \frac{1}{Z^{-1} \omega -\omega_d -i\gamma \omega}
\end{equation}
to our data for $K({\bf 0},\omega)$, where $Z, \omega_d$, and $ \gamma$ are free fitting parameters. Results from our simulations for $K({\bf 0},\omega)$ as a function of $\omega$ for $T=1.0T_c$ together with the fitted function above are shown in Fig. \ref{Fig_15}. The results for the damping parameter gamma and system parameters are in Table \ref{tab:gamma}.

\begin{widetext}
\begin{table}[htbp]
    \centering
    \begin{tabular}{  c|c|c|c||c|c|c|c} 
     \hline
        System Temp, nK & Relative Temp & Total Number $\times 10^5$ & Peak Density $\times 10^{20}m^{-3}$ & $\gamma$ & Z & $\omega_d$ (Hz) & $\omega_d \times Z (Hz)$ \\
       \hline     \hline
$357$ &$1.00$ &  1.09 & 3 &0.00376 & 0.0211 & 41.8 & 1980 \\
$385$ &$1.01$  & 1.33 &2.6 & 0.00338 & 0.0130 & 27.5 & 2117 \\
$432$ &$1.02$  &  1.79 & 2.3 & 0.00310 & 0.0132 & 26.5 & 2009\\
      \hline\hline
    \end{tabular}
    \vspace*{3mm}
    \caption{Table of simulation results.}
    \label{tab:gamma}
 \end{table}
\end{widetext}

From these results we observe that the dimensionless damping parameter increases upon lowering the temperature towards the critical one. This is expected as the damping is determined by collisions that are Bose enhanced in the degenerate regime \cite{duine2009}. We come back to this point in the discussion.

\section{Spin-current fluctuations}
\label{sec:spincurrentflucs}

In this section we consider the spin resistivity of a two-component Bose gas, and, in particular, how it is measured by determining the spin-current fluctuations in equilibrium. The spin resistivity in a cold-atom system is, because of lack of impurities and lattice vibrations, fully determined by the interactions between the two spin components of the gas. Hence, it is also referred to as the spin-drag resistivity, i.e., the friction between two spin components because of interactions, which is a concept first introduced in the context of semi-conductor spintronics \cite{scd_giovanni,weber_nature_2005}. The behavior of the spin-drag resistivity in cold atomic gases has attracted some attention recently, both for Fermi \cite{marco, bruun2011, sommer} and Bose gases \cite{duine2009, koller2012}. To directly determine the spin-drag resistivity, we consider a homogeneous two-species Bose gas above condensation temperature. We apply a spin-dependent force such that $\F_\uparrow = - \F_\downarrow \equiv \F$, and assume that the two spin species have equal density, which guarantees that there will be no net mass current. The temperature is taken to be constant, and we do not include the effects of the trapping potential here.

\subsection{Phenomenological considerations}\label{sec:3macro}

The phenomenological equations of motion for our system are \cite{duine2009}

\begin{align}
\label{eq:pheno}
n m \frac{d\v_\uparrow}{dt} &= n \F + \bgam(\v_\uparrow - \v_\downarrow), \\
n m \frac{d\v_\downarrow}{dt} &= - n \F - \bgam(\v_\uparrow - \v_\downarrow).
\end{align}
In the above, $n$ is the particle density per species, $m$ the mass of a single particle, $\v_\sigma$ the drift velocity of species with spin state $|\sigma\rangle$ ($\sigma \in \{\uparrow, \downarrow\}$, and $\bgam(\v_\uparrow - \v_\downarrow)$ an as of yet undefined function describing friction between the two spin species. We take the linear, isotropic approximation for the latter (and use the prime to denote a derivative), such that  $\Gamma_{
\alpha\beta}' \equiv d\Gamma_\alpha /dv_\beta$, and $\Gamma_{xx}' = \Gamma_{yy}' = \Gamma_{zz}' \equiv \Gamma'(0)$, and that $\bgam(\v_\uparrow - \v_\downarrow) \simeq \Gamma'(0)(\v_\uparrow - \v_\downarrow)$.
The spin current is defined by $\j_{\s} = n(\v_\uparrow - \v_\downarrow)$. In the steady state the above equations yield
\begin{align}
\j_\s = -\frac{n^2}{\Gamma'(0)} \F \equiv \sigma_\s \F \equiv \rho_\s^{-1} \F \equiv \frac{n\tau_\s}{m} \F,
\end{align}
which defines the spin resistivity $\rho_\s$ and spin conductivity $\sigma_\s$. Furthermore, $\tau_\s$ introduced above is the spin transport relaxation time. However, since the only relaxation mechanism in this set-up is due to collisions between particles with different spin that lead to spin drag, we refer to it as the spin-drag relaxation time throughout.

In linear response we have for the spin current in a homogeneous system in full generality that:
\begin{align}
\j_\s (\x,t) = \int dt \int d\x' \sigma^{(+)}_\s(\x -\x', t -t') \F(\x', t').
\end{align}
From the phenomenological equations in Eq.~(\ref{eq:pheno}) it follows that
\begin{align}
\sigma_\s^{(+)}(\k = 0, \omega) = \frac{2ni}{m}\left(\frac{1}{\omega + \frac{2i}{\tau_\s}}\right).
\end{align}
Note that $\sigma_\s^{(+)}(\k=0, 0) = \sigma_\s$, as it should. In the next section, we give another expression for the spin-drag conductivity, this time in terms of a Kubo formula that relates it to the the spin-current-spin-current correlation function.

\subsection{Kubo Formalism}

For the Kubo formalism, we start from the generic action for the system, given by

\begin{eqnarray}
\label{eq:euclaction}
&& S[\phi, \phi^*] = \int d\x \int_0^{\hbar\beta} d\tau \sum_{\sigma}
\phi^*_\sigma (\x, \tau)\nonumber \\
&& \times \left[\hbar\frac{\partial}{\partial \tau}
- \frac{\hbar^2 \bnab^2 }{2 m} - \mu\right]\phi_{\sigma}(\x,
\tau) + S_{\rm int}[\phi, \phi^*],
\end{eqnarray}
and the definition of the spin current, given by
\begin{align}
\J_s(\bx,\tau) = \frac{\hbar}{2 m i} \sum_\sigma \sigma
\left[\phi_\sigma^* (\x, \tau)) \bnab \phi_\sigma (\x, \tau) - {\rm c.c.}\right],
\end{align}
where the $\phi (\bx,\tau), \phi^* (\bx,\tau)$ are the bosonic fields associated with the creation and annihilation operators, the second term in the above expression for the spin current denotes complex conjugation, and all interactions are in $S_{\rm int}[\phi,\phi^*]$. The only important interactions for determining the spin transport properties of the gas at zero momentum are in the inter-spin $s$-wave collisions that are parameterized by the scattering length $a$. (Note that in the first part of this article this notation is used for the scattering length of a single-component gas.) To incorporate the spin-dependent force we perform the minimal substitution
\begin{align}
- i \hbar \bnab \rightarrow- i \hbar \bnab + \sigma \frac{\F}{\omega_p} e^{i \k \cdot \x - i \omega_p \tau},
\end{align}
with $\F$ the external force that leads to nonzero spin currents. A standard imaginary-time linear-response calculation now leads to the result for the spin conductivity that
\begin{align}
\sigma_\s = \frac{1}{\omega_p}\left[-\frac{\Pi(\k, i\omega_p)}{\hbar} + \frac{2n}{m}\right]~,
\end{align}
where $\Pi(\k, i\omega_n)$ is the Fourier transform of the spin-current-spin-current response function $\Pi (\bx-\bx';\tau-\tau') = \langle \J_s (\bx,\tau) \cdot \J_s (\bx',\tau') \rangle/3$, where we again assumed rotational invariance and $i\omega_n$ denotes bosonic Matsubara frequencies.
Finally, we take the $\k = 0$ element and perform a Wick rotation $i\omega_p \rightarrow \omega^{+}$ so that
\begin{align}
\sigma_\s^{(+)} (\k = 0, \omega) = \frac{1}{i\omega}\left(\frac{\Pi^{(+)}(\k = 0, \omega)}{\hbar} - \frac{2n}{m}\right),
\end{align}
where $\Pi^{(+)}(\k = 0, \omega) \equiv \Pi(\k = 0, \omega^+)$ is the retarded spin-current-spin-current correlation function at zero momentum.

Using the above results together with the phenomenological expression for the spin conductivity, we determine the retarded spin-current-spin-current correlation function in terms of the spin-transport relaxation time
\begin{align}
\nonumber \Pi^{(+)}(\k = 0, \omega) = \frac{4 n \hbar}{m} \left(\frac{i}{\omega \tau_\s + 2i}\right).
\end{align}

The equilibrium fluctuations in the spin current are characterized by the correlation function $\langle \hat \J_\s(\x, t) \hat \J_\s(\x', t')\rangle$, with $\hat \J_\s$ the spin-current expressed in terms of second-quantized Heisenberg operators. The Fourier transform of this correlation function is the so-called power spectrum $P(\k,\omega)$ for the spin-current fluctuations. We focus here on the zero-momentum power spectrum $P(\omega)\equiv P(\k=0,\omega)$.

We use the fluctuation-dissipation theorem given by
\begin{align}
P(\omega) = 2 \hbar [1 + n_B(\hbar\omega)] {\rm Im}[\Pi^{(+)}(\k = 0, \omega)],
\end{align}
to determine the power spectrum in terms of the spin-transport relaxation time. This yields
\begin{align}
P(\omega) = \left[1 + n_B (\hbar \omega) \right] \frac{2 n \hbar^2}{m} \frac{\omega \tau_\s}{\frac{\omega^2 \tau_\s^2}{4} + 1}.
\end{align}
We are now in the position to use previous results for the spin transport-relaxation time obtained by three of us \cite{duine2009}. In this latter work, this relaxation time is determined with the framework of Boltzmann transport theory. In the next section we use these results to determine the power spectrum of spin-current fluctuations.

\subsection{Results}

\begin{figure}
    \centerline{
    \mbox{\includegraphics[width=7.00in]{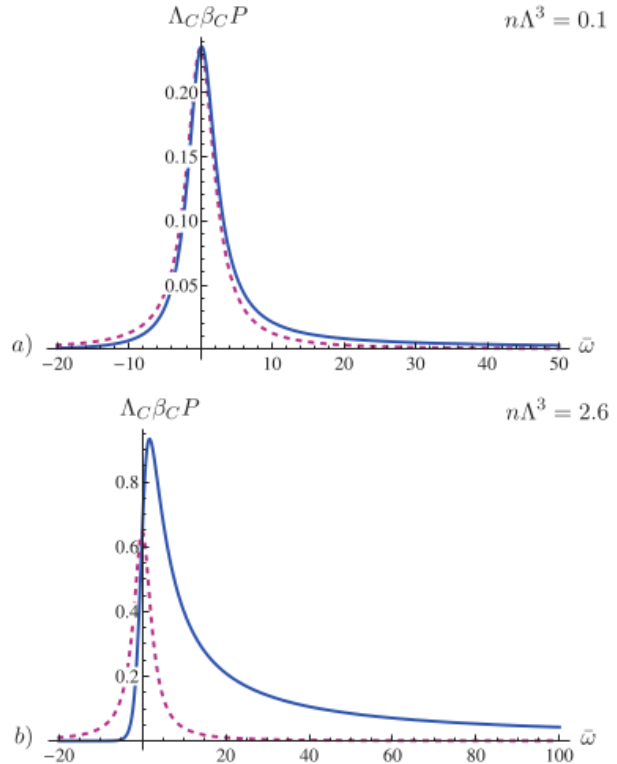}}
    }
  \caption{The dimensionless power spectrum $\Lambda_C \beta_C P(\bar{\omega})$ for $a/\Lambda_C = 0.1$ and two different limits: a) $n\Lambda^3 = 0.1$ and b) $n\Lambda^3 = 2.6$. The thick lines are the full results, the dashed ones the classical limit. }\label{fig:5.1}
\end{figure}

In Fig.~\ref{fig:5.1} the power spectrum for spin-current fluctuations is displayed as a function of dimensionless frequency $\bar{\omega} = \hbar\beta_C \omega$ with $\beta_C=1/k_B T_c$. We introduce the deBroglie wave length $\Lambda = \sqrt{2\pi\hbar^2 \beta/m}$, and its value $\Lambda_C$ at the critical temperature for Bose-Einstein condensation. We also plot the classical result which is obtained by approximating, in the fluctuation-dissipation theorem, $1+n_B (\hbar \omega) \simeq k_B T/\hbar \omega$. We note that in the low-temperature limit, the power spectrum is significantly broadened with respect to the result at higher temperature. This is a consequence of the increasing spin-transport relaxation rate $1/\tau_\s$ due to Bose-enhanced scattering at low temperatures. Also note that the classical approximation for the power spectrum ceases to be a good approximation at small temperatures.

\section{Conclusions, discussion and outlook} \label{sec:discussion}

In this article we have considered interaction effects on dynamic correlations in single and two-component Bose gases above the critical temperature for Bose-Einstein condensation. For the former, we focused on single-particle correlations, as measured in an RF spectroscopy experiment. We obtained perturbative results for a homogeneous Bose gas, and performed simulations for a trapped gas. Both results show that interactions lead to a nonzero lifetime of single-particle excitations that has significant temperature dependence.  For the case of two-component Bose gases we focused on spin-current fluctuations, and presented results on the power spectrum of these fluctuations. In this case we found broadening (in the frequency domain) of the power spectrum due to interaction effects. We note that both our simulation results for the single-particle correlation function, and the results for the power spectrum of spin-current fluctuations show decreasing lifetime upon lowering the temperature towards $T_c$. The second-order perturbation theory for the single-particle correlation function, on the other hand, shows increasing lifetime upon decreasing the temperature. These results are in agreement with the results of Ref.~\cite{rakpong}, where it was shown that over a broad temperature range the system is well described by Boltzmann transport theory which predicts Bose enhancement of scattering rates (and thus reduction of lifetime) upon approaching the critical temperature. Very close to $T_c$, i.e., in the critical region, this trend reverses and scattering rates approach zero at the critical temperature.  We hope these results will motivate future experiments to consider dynamics correlations in the non-condensed regime of Bose-Einstein condensed gases, and in particular to focus on these interesting different regimes of temperature dependence. Finally, we note that possible extensions of this work are to include the appropriate unitary-limited interaction that determines the high-frequency behavior of dynamic correlation functions, and to study other types of correlation functions.

\acknowledgements This work was supported by  the Stichting voor Fundamenteel Onderzoek der Materie (FOM), the Netherlands
Organization for Scientific Research (NWO), and by the
European Research Council (ERC) under the Seventh
Framework Program (FP7).

\end{document}